\documentclass[%
 reprint,
 amsmath,amssymb,
 aps,
]{ws-ijmpcs}

\usepackage{graphicx}
\usepackage{dcolumn}
\usepackage{bm}

\usepackage{amsmath}	
\usepackage{amssymb}	

\def\be{\begin{equation}}
\def\ee{\end{equation}}
\begin{document}

\markboth{R.O. Gomes \emph{et al.}}
{}

%
\catchline{}{}{}{}{}
%

\title{Highly magnetized neutron stars in a many-body forces formalism}

\author{R.O. Gomes}
\address{Astronomy Department, Universidade Federal do Rio Grande do Sul (UFRGS), Porto Alegre, Brazil}
\author{B. Franzon}
\address{Frankfurt Institute for Advanced Studies,
Frankfurt am Main, Germany}
\author{V. Dexheimer}
\address{Department of Physics, Kent State University, Kent OH 44242 USA}
\author{S. Schramm}
\address{Frankfurt Institute for Advanced Studies,
Frankfurt am Main, Germany}
\author{C.A.Z. Vasconcellos}
\address{Astronomy Department, Universidade Federal do Rio Grande do Sul (UFRGS), Porto Alegre, Brazil}

\maketitle

\begin{history}
\received{Day Month Year}
\revised{Day Month Year}
\published{Day Month Year}
\end{history}

\begin{abstract}
In this work, we study the effects of different magnetic field configurations in neutron stars described by a many-body forces formalism (MBF model). 
The MBF model is a relativistic mean field formalism that takes into account many-body forces by means of a meson field dependence of the nuclear interaction coupling constants. 
We choose the best parametrization of the model that reproduces nuclear matter properties at saturation and also describes massive neutron stars. 
We assume matter to be in beta-equilibrium, charge neutral and at zero temperature. 
Magnetic fields are taken into account both in the equation of state and in the structure of the stars by the self-consistent solution of the Einstein-Maxwell equations.
We assume a poloidal magnetic field distribution and  calculate its effects on neutron stars, showing its influence on the gravitational mass and deformation of the stars. 
\end{abstract}

\ccode{PACS numbers:}

\section*{Introduction}

Magnetars (neutron stars powered by their magnetic energy reservoirs) appear in nature in the form of  Soft Gamma Repeaters (SGRs) and Anomalous X-ray Pulsars (AXPs). Such objects  possess surface magnetic fields up to around $B \sim 10^{15}\,\mathrm{G}$, and are usually associated to neutron stars. Their unique environment of extreme densities and magnetic field strengths allow for the study of nuclear matter in conditions that cannot be accessed in laboratories on Earth. 

The magnetic fields in the center of magnetars are the most intense  in the universe  and, according to the Virial theorem, can reach up to $B \sim 10^{18}-10^{20}\,\mathrm{G}$  \cite{Lai1991,Cardall:2000bs,Ferrer:2010wz}. 
Magnetic effects were studied in the past in the equation of state (EoS), through Landau quantization of the particles energy levels, in different nuclear models in order to describe hyperon stars  \cite{Chakrabarty:1997ef,Broderick:2000pe,Broderick:2001qw,Sinha:2010fm,Lopes:2012nf,Casali:2013jka,Gomes:2014dka,Gao:2015jha}, quark stars  \cite{PerezMartinez:2007kw,Orsaria:2010xx,Dexheimer:2012mk,Denke:2013gha,Isayev:2015rda,Felipe:2010vr,Paulucci:2010uj} and hybrid stars \cite{Rabhi:2009ih,Dexheimer:2012qk}.

Furthermore, deformation effects must be taken into account in the modeling of such objects, as Landau quantization effects on the EoS together with the pure magnetic field contribution gives rise to a pressure anisotropy \cite{PerezMartinez:2007kw}. 
In order to calculate the macroscopic structure of highly magnetized neutron stars, one has to solve the Einstein-Maxwell coupled equations using a  non spherical metric. Such a formalism was implemented in the past by Bonazzola \emph{et al.} among others  \cite{Bonazzola:1993zz,Bocquet:1995je,Cardall:2000bs} and only recently applied to self-consistently include magnetic effects both in the EoS and structure of quark \cite{Chatterjee:2014qsa} and hybrid stars \cite{Franzon:2015sya}.

In this work, we model magnetic neutrons stars taking into account magnetic effects both in the EoS and general relativistic structure of such objects. The impact of different magnetic field configurations on global properties of magnetic neutron stars is investigated for a parametrization of the many-body forces model (MBF). First we show that the inclusion of magnetic field effects on the equation of state does not change significantly the global properties of such objects. Then the mass radius diagram is calculated for different magnetic configurations and the effects on the global properties and deformation are discussed for $2.2\,\mathrm{M_{\odot}}$ baryon mass stars. 

\section{Many-body forces Model}

The lagrangian density of the MBF model including magnetic field effects reads \cite{Gomes:2014aka}:
\small
\begin{equation}\begin{split}\label{lagrangian}
\mathcal{L}&= \underset{b}{\sum}\overline{\psi}_{b}\left[\gamma_{\mu}\left(i\partial^{\mu} -g_{\omega b}\omega^{\mu} -g_{\phi b}\phi^{\mu}
-g_{\varrho b}\mathbf{\textrm{\ensuremath{I_{3b}}\ensuremath{\varrho_3^{\mu}}}}\right)
-m^*_{b \zeta + q_{e,b}A^{\mu}}\right]\psi_{b} 
\\& +\left(\frac{1}{2}\partial_{\mu}\sigma\partial^{\mu}\sigma-m_{\sigma}^{2}\sigma^{2}\right)
+\frac{1}{2}\left(-\frac{1}{2}\omega_{\mu\nu}\omega^{\mu\nu}+m_{\omega}^{2}\omega_{\mu}\omega^{\mu}\right)
\\& +\frac{1}{2}\left(-\frac{1}{2}\boldsymbol{\varrho_{\mu\nu}.\varrho^{\mu\nu}}+m_{\varrho}^{2}\boldsymbol{\varrho_{\mu}.\varrho^{\mu}}\right)
+\left(\frac{1}{2}\partial_{\mu}\boldsymbol{\delta.}\partial^{\mu}\boldsymbol{\delta}-m_{\delta}^{2}\boldsymbol{\delta}^{2}\right)
\\& +\underset{l}{\sum}
\overline{\psi}_{l}\gamma_{\mu}\left(i\partial^{\mu}
-m_{l}\right)\psi_{l} +\underset{l}{\sum}\overline{\psi}_{l}q_{e,l}A^{\mu}\psi_{l},
\end{split}\end{equation}
\normalsize
where the indices $b$ and $l$ denote the degrees of freedom of the baryons ($p^+$, $n$) and leptons ($e^-$, $\mu^-$), respectively.
The first and last lines represents the Dirac lagrangian for baryons and leptons, respectively, while the last term describes the electromagnetic interaction by the $A^{\mu}$ field. The second line presents the lagrangian densities of the scalar-isoscalar $\sigma$ field and the vector-isoscalar $\omega$ field, responsible for the description of the attractive and repulsive features of nuclear interaction. 
The isovector fields $\delta$ (scalar) and $\varrho$ (vector) are introduced in the third line, and allow for the description of isospin asymmetry present in neutron stars.

The meson-baryon coupling appears in the first line for the vector mesons ($g_{\omega b}$, $g_{\varrho b}$ and the scalar couplings are introduced in the baryon effective masses ($m^ {*}_{\zeta b}$):
\begin{equation}
m_b^* = m_b - \left(  g^{*}_{\sigma b}\sigma + g^{*}_{\delta b}I_{3b}\delta_{3} \right),
\label{meff}
\end{equation}
The many-body forces contributions are introduced in the effective couplings of the scalar mesons as: 
\begin{equation}
g^{*}_{\sigma b} =\left(1+ \frac{g_{\sigma b}\sigma+ g_{\delta b}I_{3b}\delta_{3}}{\zeta m_{b}}  \right)^{-\zeta} g_{\sigma b},
\qquad
g^{*}_{\delta b} =\left(1+ \frac{g_{\sigma b}\sigma+ g_{\delta b}I_{3b}\delta_{3}}{\zeta m_{b}}  \right)^{-\zeta} g_{\delta b} .
\label{geff}
\end{equation}
The nonlinear contributions that arise from the scalar couplings expansion around the parameter $\zeta$ can be interpreted as higher order corrections due to the meson-meson interactions. Note that such couplings are no longer constant, depending indirectly on the density through their scalar fields contributions.

In this work, we fix $\zeta=0.040$ and use the constraints: binding energy per nucleon $B/A = -15.75\,\mathrm{MeV}$, saturation density $\rho_0 = 0.15\,\mathrm{fm^{-3}}$, symmetry energy $J_0=32\,\mathrm{MeV}$ and slope of the symmetry energy $L_0=97\,\mathrm{MeV}$. This set of parameters reproduces an effective mass of the nucleon $m^*_n = 0.66m_n$ and a compressibility $K_0=297$ (MeV) at nuclear saturation, for the couplings: $(g_{\sigma N}/m_{\sigma})^2=14.51\,\mathrm{fm^2}$, $(g_{\omega N}/m_{\omega})^2=8.74\,\mathrm{fm^2}$, $(g_{\varrho N}/m_{\varrho})^2 = 4.466\,\mathrm{fm^2}$ and   $(g_{\delta N}/m_{\delta})^2 = 0.383\,\mathrm{fm^2}$ .

The electromagnetic interaction among charged particles gives rise to a Landau quantization of their energy levels \cite{Canuto:1969cn,Broderick:2000pe,Broderick:2001qw,Gomes:2014dka}:
\begin{equation}\label{landau_levels}
e_{F}=\sqrt{\left(m_{i}\right)^{2}+k_{z}^{2}+2|q|B\nu},
\end{equation}
being $m_{i} = m_i^{*}$ in the case of baryons and the Landau quantum number $\nu$ is given in terms of the orbital and spin quantum numbers as:
\begin{equation}
\nu \equiv  l + \frac{1}{2} -  \frac{s}{2} \frac{q}{|q|}.
\end{equation}
The Landau number ranges from zero to a maximum value which avoids the particles Fermi momenta to become imaginary at zero temperature (see more details in Ref. \cite{Strickland:2012vu}).

\newpage

\section{The structure of magnetic stars}

Assuming a stationary, axi-symmetric
spacetime and the Maximum-Slice-Quasi-Isotropic coordinates (MSQI), the  line element in the 3+1 decomposition of space-time is:
\be
ds^2 = -N^{2}dt^{2} + A^2(dr^2 + r^2 d\theta^2) + B^{2}r^{2}\sin^{2}\theta( d\phi - \omega dt)^{2},
\label{metric}
\ee 
where $N(r,\theta)$, $A(r,\theta)$, $B(r,\theta)$ and $\omega(r, \theta)$ are functions only of the coordinates $(r, \theta)$. 

Introducing such a metric in the Einstein's field equations, one obtains the following equation system for the metric potentials:
\be 
\Delta_{2} [(NB-1) r \sin \theta] = 8\pi NA^2 B r \sin \theta (S^{r}_{r} + S^{\theta}_{\theta}),
\label{Bfinal}
\ee
\be 
\Delta_{2} [{\rm{ln}} A + \nu] = 8\pi A^2 S^{\phi}_{\phi} + \frac{3 B^2 r^2 \sin^2 \theta}{4 N^2} \partial \omega \partial \omega - \partial \nu \partial \nu,
\label{Afinal}
\ee
\be 
\Delta_{3} \nu = 4\pi A^2 (E + S) + \frac{B^2 r^2 \sin^2 \theta}{2N^2} \partial \omega \partial \omega - \partial \nu \partial( \nu + {\rm{ln}} B),
\label{Nfinal}
\ee
\be
\left[ \Delta_{3} - \frac{1}{r^2 \sin^2 \theta} \right] (\omega r \sin \theta) = -16\pi \frac{NA^2}{B^2} \frac{p_{\phi}}{r \sin \theta} 
 + r \sin \theta  \partial \omega \partial(\nu - 3 {\rm{ln}} B),
\ee
following the notation:
\begin{align}
& \Delta_{2} = \frac{ \partial^2}{\partial r^2} + \frac{1}{r}\frac{ \partial}{\partial r} + \frac{1}{r^2}\frac{ \partial^2}{\partial \theta^2}, \qquad  \nu = {\rm{ln}} N, \\
& \Delta_{3} = \frac{ \partial^2}{\partial r^2} + \frac{2}{r}\frac{ \partial}{\partial r} + \frac{1}{r^2}\frac{ \partial^2}{\partial \theta^2} + \frac{1}{r^2 \tan \theta}\frac{ \partial}{\partial \theta}, \\
& \partial \omega \partial \omega := \frac{\partial \omega}{\partial \omega}\frac{\partial \omega}{\partial r} + \frac{1}{r^2}\frac{\partial \omega}{\partial \theta}\frac{\partial \omega}{\partial \theta}. 
\end{align}

In order to include magnetic field effects on the structure of the stars, we solve the coupled Einstein-Maxwell field equations system. The energy $E$, total momentum density flux $J_{\phi}$ and total stress tensor $S$ are decomposed in perfect fluid (PF) and  electromagnetic (EM) contributions.

We assume a poloidal magnetic field distribution. In this case, the electric and magnetic field components (measured by the Eulerian observer $\mathcal{O}_{0}$) are written, respectively, as \cite{Bocquet:1995je}:
\be
E_{\alpha} = \left( 0 , \frac{1}{N} \left[  \frac{\partial A_{t}}{\partial r} + \omega \frac{\partial A_{\phi}}{\partial r}\right ] , \frac{1}{N} \left[  \frac{\partial A_{t}}{\partial \theta} + \omega \frac{\partial A_{\phi}}{\partial \theta}\right ]   , 0 \right),
\ee
\be
\hspace{-2cm} B_{\alpha} = \left( 0 , \frac{1}{B r^{2} \sin \theta} \frac{\partial A_{\phi}}{\partial \theta}, - \frac{1}{B \sin \theta} \frac{\partial A_{\phi}}{\partial r} , 0  \right),
\ee
where $A_{t}$ and $A_{\phi}$ are the two non-zero components of the electromagnetic four-potential $A_{\mu} = (A_{t}, 0 , 0 , A_{\phi})$.

The Faraday tensor $F_{\mu\nu}$ makes some of the Maxwell equations automatically satisfied, remaining only the Gauss and Ampere equations to  be solved. The Maxwell-Gauss equation for $A_{t}$ reads:
\begin{align} 
\Delta_{3}A_{t} = &- \mu_{0} A^2 (g_{tt}j^{t} + g_{t\phi}j^{\phi}) 
- \frac{B^2}{N^2}\omega r^2 \sin^2 \theta \partial A_{t} \partial N^{\phi} \nonumber \\
&-\left(1+\frac{B^2}{N^2} r^2 \sin^2 \theta \omega^2 \right) \partial A_{\phi}\partial \omega -(\partial A_{t} + 2 \omega \partial A_{\phi}) \partial ({\rm{ln}}B - \nu) \nonumber \\
&-2\frac{\omega}{r} \left( \frac{\partial A_{\phi}}{\partial r} + \frac{1}{r \tan\theta} \frac{\partial A_{\phi}}{\partial r}\right),
\label{maxwellgauss}
\end{align}
and from the Maxwell-Ampere equation, we have an equation for $A_{\phi}$:
\begin{align}
&\left[\Delta_{3} -\frac{1}{r^2 \sin^2 \theta}\right] \left( \frac{A_{\phi}}{r \sin \theta}\right) = - \mu_{0} A^2 B^2 ( j^{\phi} - \omega j^{t})r \sin \theta \nonumber \\
& - \frac{B^2}{N^2} r \sin \theta \partial \omega (\partial A_{t}+\omega \partial A_{\phi}) + \frac{1}{r} \partial A_{\phi} \partial ({\rm{ln}}B - \nu).
\label{maxwellampere}
\end{align}

In this approach, the equation of motion reads:
\be 
H \left(r, \theta \right) + \nu \left(r, \theta \right) + M \left(r, \theta \right) = const ,
\ee
where $H(r,\theta)$ is the heat function defined in terms of the baryon number density $n$: 
\be H =
\int^{n}_{0}\frac{1}{e(n_{1})+p(n_{1})}\frac{d P}{dn}(n_{1})dn_{1}.
\label{heat}
\ee
The magnetic potential $M(r,\theta)$ is given by:
\be M \left(r, \theta \right) = M \left(
A_{\phi} \left(r, \theta \right) \right) \equiv -
\int^{0}_{A_{\phi}\left(r, \theta \right)} f\left(x\right)
\mathrm{d}x, 
\ee
with  $f(x)$ being the current function, which we choose to be constant in this work.  According to Ref.~\cite{Bocquet:1995je}, other choices for $f(x)$ are possible, but they do not change the general conclusions.
Different values of the current function generate different currents and, consequently, different magnetic fields strenghs distributions throughout  stars. For more details of this formalism, see Refs. \cite{Bocquet:1995je,Chatterjee:2014qsa,Franzon:2015sya} and references therein.

\section{Results and Discussion}

We use the EoS of the MBF model as an input to  calculate the strucuture of the magnetic stars, following the formalism presented in the last section. Note that such calculations are self-consistent, as we are taking into account magnetic effects both on the EoS and on the star structure. 

\begin{figure}
\centering
   \includegraphics[width=1.03\linewidth]{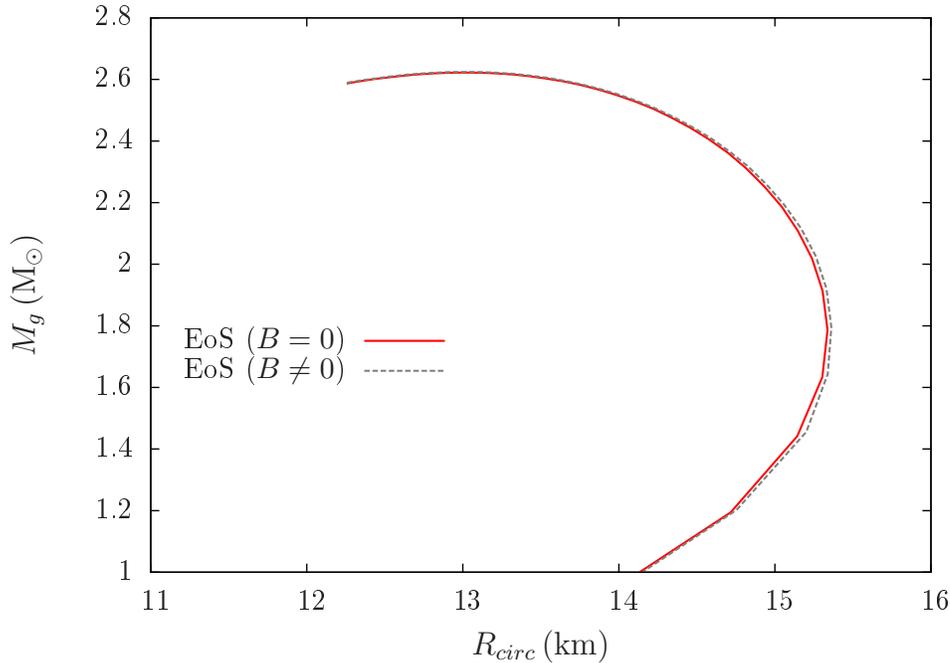}
    \caption{Mass-radius diagram for the parametrization $\zeta=0.040$ of the MBF model with a current function $j_0 = 3.5\times 10^{15} \mathrm{A/m^2}$ including and not including magnetic effects on the EoS. The vertical axis shows the gravitational mass and the horizonal one the equatorial radius.}
\label{mg}
 \end{figure}

The first important result we obtain is shown in Figure \ref{mg}, in which the mass-radius diagram is presented for magnetic nucleonic stars. The color dashed line shows the results obtained by including magnetic effects both in the EoS and in the strucuture of the stars. The full red line shows the results obtained by not taking the Landau quantization effects into account on the EoS (only in the structure of the stars). As one can see, for a fixed value of magnetic moment of the stars, including magnetic effects on the EoS generates no significant impact on the gravitational mass and radius of magnetic stars. This indicates that magnetic effects on magnetars are only significant on the macroscopic scale i.e., on the structure of such stars. These results agree with the previous ones in the literature for magnetic quark \cite{Chatterjee:2014qsa} and hybrid stars \cite{Franzon:2015sya}.

\begin{figure}
\centering
   \includegraphics[width=1.03\linewidth]{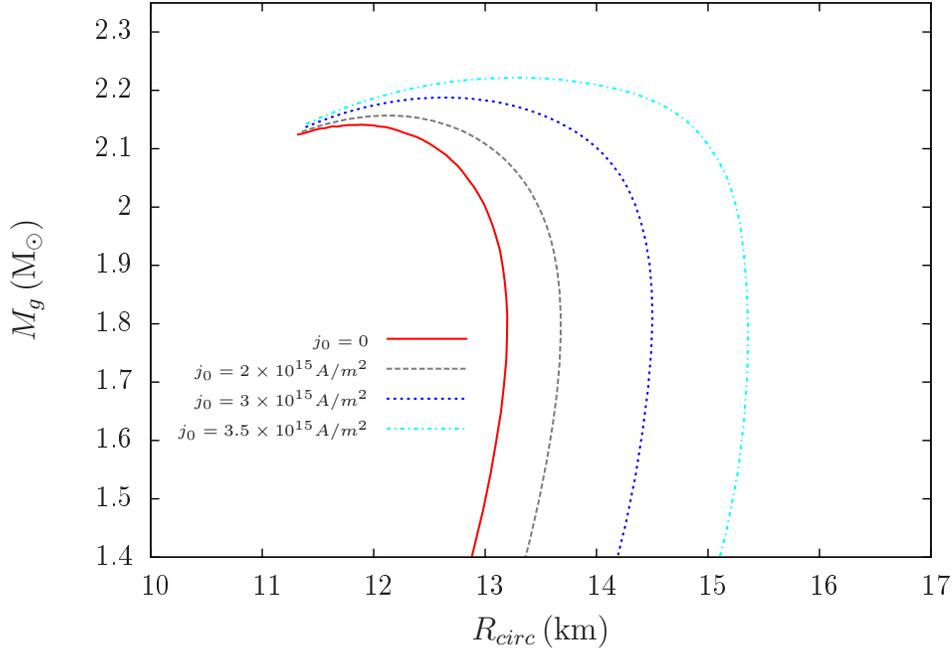}
    \caption{Same diagram as Figure \ref{mg}, but for different current functions.}
\label{mr}
 \end{figure}

Figure \ref{mr} shows the mass-equatorial radius diagram for different current function values. As already mentioned in last section, different choices of the current function alter the magnetic field  strenghs throughout stars. Higher current values allow for stronger magnetic fields inside stars. 
The red full curve represents the non magnetic case, while the dashed lines include non zero values for the current function. 

Strong magnetic fields have direct impact on the global properties of stars. Since the Lorenz force opposes gravity, the extra magnetic energy is responsible for increasing the gravitational mass of the stars, as well as their radius. 

\begin{figure}
\centering
   \includegraphics[width=1.03\linewidth]{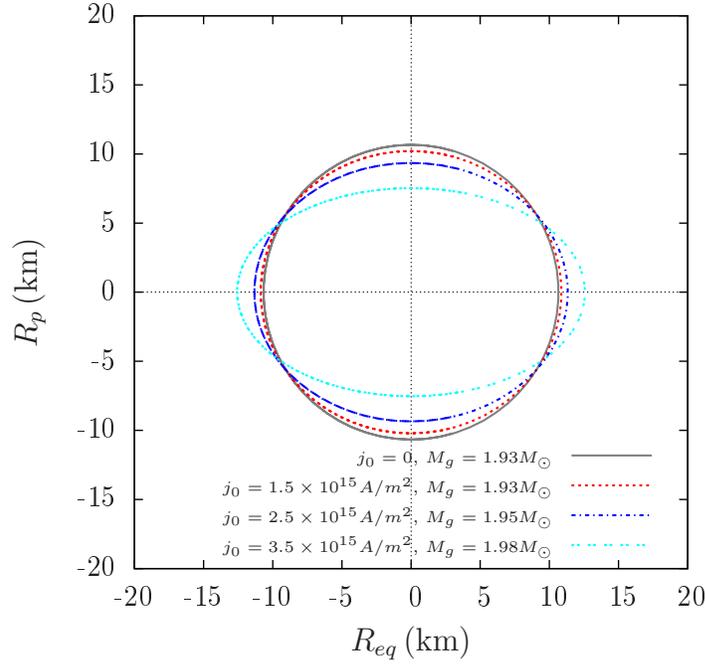}
    \caption{Cross section profile of several $M_b = 2.2\,\mathrm{M_{\odot}}$ star for different magnetic field configurations. The vertical and horizontal  axes show, respectively, the polar and equatorial radii. }
\label{deformation}
 \end{figure}

Moreover, as a poloidal magnetic field distribution is assumed, stars that endow stronger magnetic fields become more oblate and, consequently, more deformed. This effect can be seen in Figure \ref{deformation}, where we show the cross section of several $M_b=2.2\,\mathrm{M_{\odot}}$ stars. The surface and central magnetic fields of these stars are shown in Table \ref{bfields}. Our results show that the deformation of magnetic stars with such profiles are extremely high, reaching $r_{eq} =1.7\, r_p$ for the star with current function $j_0 = 3.5\times 10^{15} \mathrm{A/m^2}$.

Several works in the past have included magnetic fields only on the equation of state of matter inside neutron stars, ignoring the deformation of the stars and solving the spherical TOV equations. In this work we have applied the MBF model to describe nucleonic matter inside neutron stars taking magnetic effects into account both on the EoS and on the structure of the stars. Our results show that, different from what was concluded in the past, only the magnetic field effects on the structure of stars are signigicant for the determination of the gravitational mass and radii (polar and equatorial) of stars. We also emphasize the neglecting the deformation of stars in the calculations which involve strong magnetic fields ($B\sim 10^{17}\,\mathrm{G}$) leads to considerable error on the determination of these quantities.
Finally, we mention that the increase of the radius of stars due to the extra repulsion from magnetic fields decreases the central density of stars, having dramatic impact on the population of such objects \cite{Franzon:2015sya}. The study of the population of magnetic stars is going to be  investigated in detail within the MBF model in a future work.

\begin{table}
  \caption{Characteristics of magnetized nucleonic stars with $M_b=2.2\,\mathrm{M_{\odot}}$ for different current function values.}
 \label{bfields}
\begin{center}
\begin{tabular}{ccc}
 \hline
$j_0\,(10^{15}\mathrm{A/m^2})$ &  $ B_s\,(10^{17}\mathrm{G}) $ & $ B_c\,(10^{17}\mathrm{G}) $  \\
  \hline \hline
 $1.0$ &  $1.03$  & $3.48$  \tabularnewline
 $2.0$ &  $1.47$  & $4.66$  \tabularnewline
 $3.0$ &  $2.70$  & $7.03$  \tabularnewline
 $3.5$ &  $3.80$  & $8.09$  \tabularnewline
  \hline\hline
  
  \end{tabular}
\end{center}
\end{table}


\end{document}